\documentclass[11pt]{article}

\usepackage{fullpage}
\usepackage{setspace}
\usepackage{amsthm}
\usepackage{amsmath, amssymb}
\usepackage[boxed]{algorithm}
\usepackage{algorithmic}
\usepackage{enumerate}
\usepackage{graphicx}
\usepackage{color}

\newtheorem{theorem}{Theorem}
\newtheorem{lemma}[theorem]{Lemma}

\newtheorem{definition}[theorem]{Definition}

\newcommand{\set}[1]{\left\{#1\right\}}
\newcommand{\card}[1]{\left|#1\right|}
\newcommand{\floor}[1]{\left\lfloor#1\right\rfloor}
\newcommand{\ceil}[1]{\left\lceil#1\right\rceil}

\newcommand{\mI}{\mathcal I}
\newcommand{\mA}{\mathcal A}
\newcommand{\mU}{\mathcal U}
\newcommand{\mT}{\mathcal T}
\newcommand{\mS}{\mathcal S}
\newcommand{\mC}{\mathcal C}
\newcommand{\mF}{\mathcal F}
\newcommand{\mV}{\mathcal V}
\newcommand{\mD}{\mathcal D}
\newcommand{\mB}{\mathcal B}

\newcommand{\OPT}{\mathsf{OPT}}
\newcommand{\opt}{\mathsf{opt}}
\newcommand{\cost}{\mathsf{cost}}
\newcommand{\CBall}{\mathsf{CBall}}
\newcommand{\FBall}{\mathsf{FBall}}

\usepackage{boxedminipage}

\usepackage{eqparbox,array}

\newtheorem{claim}[theorem]{Claim}
\newenvironment{proofclaim}{\begin{trivlist}
    \item[\hskip\labelsep {\it Proof of Claim}.]}{\QED \end{trivlist}}
\newcommand{\QED}{\hfill $\square$}

\def \imagedir{.}
\def \imagewidth{0.6\textwidth}

\begin{document}
\title{Approximating $k$-Median via Pseudo-Approximation}
\author{
Shi Li \\ Princeton, USA  \\  shili@cs.princeton.edu
\and 
Ola Svensson \\ EPFL, Switzerland \\  ola.svensson@epfl.ch
}
\maketitle 

\begin{abstract}
 We present a novel approximation algorithm for $k$-median that achieves   an approximation guarantee of
 $1+\sqrt{3}+\epsilon$, improving upon the decade-old ratio of $3+\epsilon$. Our approach is based
 on two components, each of which, we believe, is of independent interest.

First, we show that in order to give an $\alpha$-approximation algorithm for $k$-median, it is
sufficient to give a \emph{pseudo-approximation algorithm} that finds an $\alpha$-approximate
solution by  opening $k+O(1)$
facilities. This is a rather surprising result as there exist instances for which opening $k+1$
facilities may lead to  a significant  smaller cost than if only $k$ facilities were opened.
 
Second, we  give such  a pseudo-approximation algorithm with $\alpha=
1+\sqrt{3}+\epsilon$. 
Prior to our work, it was not even known whether opening $k + o(k)$ facilities would help improve the approximation ratio.

\end{abstract}

\section{Introduction}

Suppose you wish to select $k$ polling stations for the US election so as to minimize the average
distance each voter has to travel to his/her closest polling station. Then you need to solve the
classic $\textsc{NP}$-hard $k$-median problem that we shall design better approximation algorithms
for in this paper. Formally, a $k$-median instance $\mI$ is defined by the tuple $(k, \mF,\mC, d)$,
where $k$ is the number of facilities allowed to be opened, $\mF$ is a set of potential facility
locations, $\mC$ is a set of clients, and $d$ is a distance metric over $\mF \cup \mC$. The goal is
to open a set $\mS \subseteq \mF$ of $k$ facilities so as to minimize $cost_{\mI}(\mS)  = \sum_{j
  \in \mC} d(j, \mS)$, where $d(j, \mS)$ denotes the distance from $j$ to its nearest facility in $\mS$. When
$\mF = \mC = X$, a solution $\mS$ partitions the set of points into what is known as  clusters and
thus the objective measures how well $X$ can be partitioned into $k$ clusters. The $k$-median problem has
numerous applications, starting from clustering and data mining \cite{BFM98}  to assigning efficient
sources of supplies to minimize the transportation cost(\cite{KH63,Man64}).

The difficulty of the $k$-median problem lies in the hard constraint that only $k$ facilities are
allowed to be opened. Indeed, without such a constraint, we could simply open all facilities. Early
approaches~\cite{LV92A,LV92B,KPR98} overcame this difficulty by giving pseudo-approximations that
obtain better guarantees while violating the mentioned constraint by opening $k + \Omega(k)$
facilities. The first constant factor approximation algorithm that opens $k$ facilities is due to Charikar et al.\ \cite{CGT99}. Based
on LP rounding, their algorithm produces a $6\frac23$-approximation. Several of the
ideas in~\cite{CGT99} are inspired from constant factor approximation algorithms obtained for the closely related metric
uncapacitated facility  location (UFL) problem. The UFL problem has  similar input as $k$-median
but instead of giving an upper bound $k$ on the number of facilities we can open, it
specifies an opening cost $f_i$ for each facility $i \in \mF$. The goal is to open a set of
facilities $\mS$ that minimizes the sum of the opening costs and connection costs, i.e.,
$\sum_{i\in
 \mS}f_i + \cost_{\mI}(\mS)$.

The connection between UFL and $k$-median is motivated by basic economic theory: if we let the  opening
costs of facilities be small then a ``good'' solution to UFL will open many facilities whereas if we let the
opening costs of facilities be  large   then a good solution will only open  few facilities.
By appropriately selecting the cost of facilities, one can therefore expect that an algorithm for
UFL opens close to $k$ facilities and therefore almost also gives a solution to the $k$-median
problem. This is the intuition of the concept of bi-point solutions that we define in
Section~\ref{sec:prelim}. Jain and Vazirani first exploited this concept in a beautiful paper~\cite{JV01}
to obtain a $6$-approximation algorithm for $k$-median using their  $3$-approximation
primal-dual algorithm for UFL.
The factor $3$ was later improved by Jain et al.~\cite{JMS02} to $2$ resulting in a
$4$-approximation algorithm for $k$-median.

In spite of the apparent similarities between UFL and $k$-median, current techniques give a
considerable better understanding of the approximability of UFL. For UFL and its variants, there has
indeed been a steady stream of papers giving improved
algorithms~\cite{LV92B,STA97,JV01,CS04,KPR98,CG99,JMM03,JMS02,MYZ06,Byr07}. The current best
approximation algorithm is due to Li~\cite{Li11}. He combined an algorithm by Byrka~\cite{Byr07} and
an algorithm by Jain et al~\cite{JMS02} to achieve an approximation guarantee of $1.488$. This is
close to being best possible, as it is hard to approximate UFL within a factor of $1.463$~\cite{GK98}. In contrast
there has been less progress for $k$-median and the approximability gap is larger. The best known
approximation algorithm is the local search algorithm given by Arya et al.~\cite{AGK01}. They showed
that if there is a solution $\mF'$, where any $p$ swaps of the open facilities cannot improve the
solution, then $\mF'$ is a $3 + 2/p$ approximation. This leads to a $3 + \epsilon$ approximation
that runs in time $n^{2/\epsilon}$. On the negative side, Jain et al.~\cite{JMS02} proved that the
$k$-median problem is hard to approximate within a factor $1+2/e\approx 1.736$. Moreover, the
natural linear programming relaxation of $k$-median is known to have an integrality gap of at least
$2$. The best upper bound is by Archer et al.~\cite{ARS03}, who showed that the integrality gap is
at most 3 by giving an exponential time rounding algorithm that requires to solve the maximum
independent set problem.  

As alluded to above, the main difficulty of the $k$-median problem is the
hard constraint that we can open at most $k$ facilities. In this paper we take a different approach
that allows us to relax this constraint and thereby addressing the problem from a novel point of
view using what we call a pseudo-approximation algorithm. This leads to the improved approximation
algorithm breaking the barrier of $3$ that we discuss next.

\subsection{Our Results}

Our improved approximation algorithm can be stated as follows.
\begin{theorem}
\label{theorem:main}
There is an algorithm which, given a $k$-median instance $\mI$ and a number $\epsilon > 0$, produces a $1 + \sqrt{3} + \epsilon$-approximate solution to $\mI$ in running time $O\left(n^{O(1/\epsilon^2)}\right)$.
\end{theorem}

Our algorithm contains two main components, each of which, we believe, is of independent
interest. First, we show that in order to give an approximation algorithm for $k$-median, it
suffices to give a \emph{pseudo-approximation algorithm} $\mA$ which, given a $k$-median instance
$\mI$, outputs a set $\mS \subseteq \mF$ of $k+c$ facilities with $\cost_{\mI}(\mS) \leq \alpha
\opt_{\mI}$, where $\opt_{\mI}$ is the cost of optimum solution for $\mI$.
Given such an algorithm $\mA$ as a black box, we can design an $\alpha+\epsilon$-approximation
algorithm $\mA'$ whose running time is $n^{O(c/\epsilon)}$ times that of $\mA$. Interestingly, the
instance (see Figure~\ref{fig:gap-2-instance}) that gives the integrality gap of $2$ for the natural LP relaxation of $k$-median  vanishes if
we allow the integral solution to open $k+1$ facilities.  This suggests that our reduction may bring
in new avenues for approximating $k$-median. In particular, we find the following open problem 
interesting: given a $k$-median instance $\mI$, what is the maximum ratio between the cost of the
optimum integral solution of $\mI$ with $k+1$ open facilities, and the LP value (with $k$ open
facilities)?  

To complement the first component, we give the aforementioned pseudo-approximation algorithm $\mA$
with $\alpha = 1+\sqrt{3}+\epsilon$.  Prior to our work, it was not even known whether opening $k +
o(k)$ facilities would help improve the approximation ratio; all known pseudo-approximation
algorithms require $k + \Omega(k)$ open facilities.  In contrast, our algorithm only opens $k +
O(1/\epsilon)$ facilities. The algorithm $\mA$ contains 2 steps. We obtain a \emph{bi-point solution} for
$k$-median using the algorithm of \cite{JMS02}. We lose a factor of $2$ in this step. Then, we
convert the bi-point solution into an integral solution with $k+O(1/\epsilon)$ open facilities, losing
another factor of $\frac{1+\sqrt{3}+\epsilon}{2}$ in the approximation ratio.  We remark that  if we had insisted on
opening $k$ facilities, then a factor of $2$ has to be lost in the last step as the instance
achieving an integrality gap of $2$ has  a bi-point solution.

Theorem~\ref{theorem:main} does not give a better upper bound on the integrality gap of the natural LP due to the
following reason: instead of running the pseudo-approximation algorithm $\mA$ on the input instance
$\mI$, we run it on a residual instance $\mI'$ obtained from $\mI$ by removing a subset of
facilities that the optimal solution does not  open.  The way we obtain $\mI'$ is to guess $O(1/\epsilon^2)$
``events'' and let $\mI'$ be the instance conditioned on these events.  Due to this nature, our
algorithm can be converted to a rounding algorithm based on solving an $O(1/\epsilon^2)$-level LP in
the Sherali-Adams hierarchy.  Instead of guessing the $O(1/\epsilon^2)$ events, we can now find these
events explicitly by looking at the LP solution. Conditioning on these events, we obtain a
fractional solution of the basic LP. By rounding this LP, we obtain a
$1+\sqrt{3}+\epsilon$-approximate solution. Thus, our approach can be seen to give an
$1+\sqrt{3}+\epsilon$-upper bound on the integrality gap of the $O(1/\epsilon^2)$-level LP in the
Sherali-Adams hierarchy. Our result was in fact first obtained by studying the power of the
Sherali-Adams hierarchy for the $k$-median problem. However, as it can also be obtained using a
combinatorial approach with less
cumbersome notation, we have chosen to present that approach.

\subsection{Preliminaries}
\label{sec:prelim}

Given a $k$-median instance $\mI = (k, \mF, \mC, d)$, a \emph{pseudo-solution} to $\mI$ is a
set $\mS\subseteq \mF$. A pseudo-solution $\mS$ satisfying
$\card{\mS}\leq k$ is a \emph{solution} to $\mI$; a pseudo-solution $\mS$ with $\card{\mS} \leq k +
c$, for some number $c \geq 0$, is called a \emph{$c$-additive (pseudo-)solution}.
The cost of a pseudo-solution $\mS$ to $\mI$ is defined as $\cost_\mI(\mS) = \sum_{j\in \mC} d(j, \mS)$, where $d(j, \mS)$ denotes the distance from $j$ to its closest facility in $\mS$.  We let $\OPT_\mI$ denote an optimal solution to
$\mI$, i.e., one of minimum cost,  and we let $\opt_\mI = \cost_\mI(\OPT_\mI)$. 
To avoid confusion we will throughout the paper
assume that the optimal solution is unique and that the concept of closest facility (or client) is
also uniquely defined. This can be achieved either by slightly perturbing the metric or by simply breaking ties in an arbitrary but fixed way.

When considering
a client or facility, it shall be convenient to argue about close clients or
facilities. For any $p \in \mF \cup \mC$ and $r \geq 0$, we therefore define
$\FBall_{\mI}(p, r) = \set{i \in \mF : d(p, i) < r}$ and $\CBall_{\mI}(p, r) = \set{j \in \mC : d(p,
  j) < r}$ to be the set of facilities and clients within distance less than $r$ from $p$,
respectively.  When $\mI$ is clear from the context, we omit the subscripts in $\cost_{\mI}$,
$\OPT_\mI$,  $\opt_\mI$, $\FBall_{\mI},$ and $\CBall_{\mI}$.



The  standard linear programming relaxation for the $k$-median problem is formulated as follows.
\begin{subequations}
\begin{alignat}{3}
\nonumber & \textrm{minimize} & \textstyle \sum_{i\in \mF, j\in \mC} d(i,j) x_{ij}  && \\[2mm]
\label{eq:kmedian} &\textrm{subject to}&  \textstyle \sum_{i\in \mF} y_i   &  \leq     k & \\[2mm]
\label{eq:connect}&& \textstyle \sum_{i\in \mF} x_{ij}  & =       1& \qquad  j\in \mC \\[2mm]
\label{eq:copen}&& \textstyle  x_{ij}               & \leq    y_i& \qquad i\in \mF, j \in \mC\\[2mm]
\label{eq:binary}&& \textstyle  x_{ij}, y_i & \in    [0,1]& \qquad i\in \mF, j \in \mC 
\end{alignat}
\end{subequations}
Constraint~\eqref{eq:kmedian} says that we are allowed to open at most $k$ facilities,
Constraint~\eqref{eq:connect} says that we must connect each client, and Constraint~\eqref{eq:copen}
says that if we connect a client to a facility then that facility has to be opened.

\begin{figure}
\centering
\includegraphics[width=\imagewidth]{\imagedir 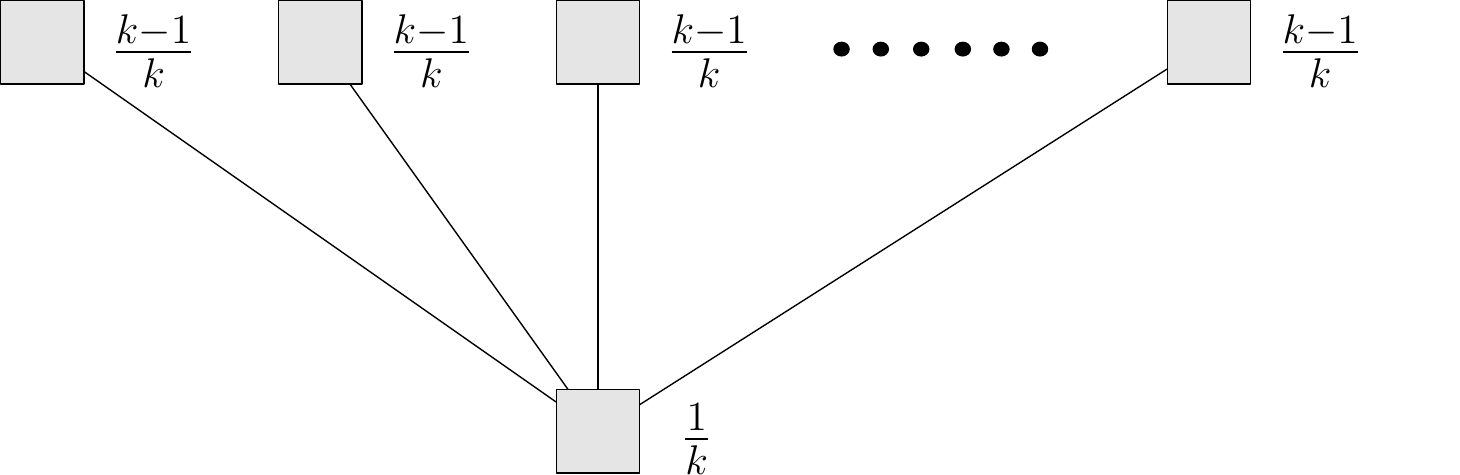}
\caption{Instance that gives integrality gap $2/(1+1/k)$ and the optimal fractional solution. We have $k+2$ facilities and $k+1$ clients co-located with the top $k+1$ facilities. All edges in the graph have length 1. The optimal integral solution has cost 2, while the optimal fractional solution has cost $(k+1)\left(\frac{k-1}{k}\cdot 0 + \frac{1}{k}\cdot 1\right) = 1+1/k$.} 
\label{fig:gap-2-instance}
\end{figure}

As mentioned earlier, the above linear programming has an integrality gap of $2$, even when the
underlying metric is a tree. The instance that gives the integrality gap of $2$ is depicted in
Figure~\ref{fig:gap-2-instance}. It is a star with $k+1$ leaves. The center of the star is a
facility and the leaves are both facilities and clients. Note that a pseudo-solution that opens all
leaves, i.e., $k+1$ facilities, has cost $0$ whereas any solution that opens only $k$ facilities
has cost $2$. The solution to the linear program obtained by a linear combination of the
pseudo-solution that opens all leaves and the solution that only opens the center of the star has
cost $1+1/k$ yielding the integrality gap of $2$ when $k$ tends to infinity. In general, a solution that  is a
linear combination of two pseudo-solutions is called a \emph{bi-point (fractional) solution}. As this
concept is important for our pseudo-approximation algorithm, we state its formal definition.

\begin{definition}[bi-point (fractional) solution]
\label{def:bi-point-solution}
Let $\mI = (k, \mF, \mC, d)$ be a $k$-median instance. Let $\mS_1$ and $\mS_2$ be two pseudo-solutions to $\mI$ such that $\card{\mS_1} \leq k  < \card{\mS_2}$. Let $a \geq 0, b \geq 0$ be the real numbers such that $a + b = 1$ and $a\card{\mS_1} + b\card{\mS_2} = k$.  Then, the following fractional solution to $\mI$, denoted by $a\mS_1 + b\mS_2$,  is called a bi-point (fractional) solution:
\begin{enumerate}
\item $y_i = a1_{i \in \mS_1} + b1_{i \in \mS_2}$;
\item $x_{i, j} = a1_{\mathsf{clst}(i, \mS_1, j)} + b1_{\mathsf{clst}(i, \mS_2, j)}$, where $\mathsf{clst}(i, \mS, j)$ denotes the event that $i$ is the closest facility in $\mS$ to $j$.
\end{enumerate}
\end{definition}
It is easy to see that the cost of the fractional solution $a\mS_1 + b\mS_2$ is exactly $a\cost_\mI(\mS_1) + b\cost_\mI(\mS_2)$. 
Jain and Vazirani~\cite{JV01} gave a \emph{Lagrangian multiplier preserving} $3$-approximation for
UFL, which  immediately yields an algorithm which produces a bi-point solution
whose cost is at most $3$ times the optimum. Together with an algorithm which converts a
bi-point solution to an integral solution at the cost of a factor $2$, \cite{JV01} gave a
$6$-approximation for $k$-median. Later, the factor $3$ was improved by Jain et al.\ \cite{JMS02} to
$2$. We now formally state the result of \cite{JMS02}.

\begin{theorem}[\cite{JMS02}]
\label{theorem:construct-bi-point-solution}
Given a $k$-median instance $\mI$, we can find in polynomial time a bi-point solution $a\mS_1 +
b\mS_2$ to $\mI$ whose cost is at most 2 times the cost of an optimal solution to $\mI$. 
\end{theorem}

\subsection{Overview of the Algorithm}
\ifdefined \stoc
\begin{algorithm*}[ht!]
\caption{Enumeration of $k$-median instances.}
\label{algo:preprocess}
 \algsetup{indent=1cm}
\begin{algorithmic}
\REQUIRE{a $k$-median instance $\mI = (k, \mF, \mC, d)$ and a positive integer $t$}\;
\ENSURE{a set of $k$-median instances so that at least one satisfies the properties of Lemma~\ref{lemma:sparseenum}}\;\\[2mm]
\STATE \textbf{for all} $t'\leq t$
facility-pairs $(i_1, i_1'), (i_2, i_2'), \dots, (i_{t'}, i_{t'}')$ \textbf{output} $(k, \mF',
\mC, d)$, where \; \\[4mm]
\STATE \hspace{\algorithmicindent} $\mF' = \mF \setminus \bigcup_{z=1}^{t'}\FBall(i_z, d(i_z,i'_z))$
\COMMENT{{\scriptsize the facilities that are  closer to $i_z$ than $i'_z$ is to $i_z$ are removed \quad \ \ \ \ \ \ \ \ \ \ \ \ \ \ \ \ \ \  }}\;\\[2mm]
\end{algorithmic}
\end{algorithm*}
\fi

The two components of our algorithm are formally stated in Theorem~\ref{theorem:transformation} and
Theorem~\ref{theorem:pseudo-approximation}, whose proofs will be given in
Sections~\ref{section:transformation} and~\ref{section:pseudo-approximation},
respectively. Together they immediately  imply Theorem~\ref{theorem:main}.
\begin{theorem}
\label{theorem:transformation}
Let $\mA$ be a $c$-additive $\alpha$-approximation algorithm for $k$-median, for some $\alpha >
1$. Then, for every $\epsilon >0$  there is a $\alpha+\epsilon$-approximation algorithm $\mA'$ for $k$-median whose running time is $O\left(n^{O(c/\epsilon)}\right)$ times the running time of $\mA$.
\end{theorem}
\begin{theorem}
\label{theorem:pseudo-approximation}
There exists a polynomial time algorithm which, given a $k$-median instance $\mI = (k, \mF,  \mC,
d)$ and $\epsilon > 0$, produces an $O(1/\epsilon)$-additive $1+\sqrt{3}+\epsilon$-approximate
solution to $\mI$. 
\end{theorem}

We now provide more details about the proof of the two theorems.  At first glance, it seems that the
transformation from a pseudo-approximation to a real approximation stated in
Theorem~\ref{theorem:transformation} is impossible, since there are cases where allowing $k+1$ open
facilities would give much smaller cost than only allowing $k$ open facilities. However, we show that we can
pre-process the input instance so as to avoid these problematic instances.  Roughly speaking, we
say that a facility $i$ is dense if the clients in a small ball around $i$ contribute a lot to the
cost of the
optimum solution $\OPT$ (see Definition~\ref{def:sparseinstance}). We guess the $O(1/\epsilon)$ densest facilities and their respective nearest
open facilities in $\OPT$.  Then for each such dense facility $i$ whose nearest open facility in $\OPT$ is
$i'$, we remove all facilities that are closer to $i$ than $i'$ (including the dense facility
$i$). 
 Then we get a residual instance in which the gap between the costs of
opening $k + O(1)$ and $k$ facilities is small. The pseudo-approximation
algorithm is then applied to this residual instance. 

For example, consider the integrality gap instance depicted in Figure~\ref{fig:gap-2-instance} and
let $\OPT$ be the optimal solution that opens the center and $k-1$ leaves. Then the two leaves that
were not opened contribute a large fraction of the total cost (each contributes $\opt/2$ to be precise) and the
two corresponding facilities are dense. By removing these dense facilities in a
preprocessing step, the gap between the costs of opening $k+O(1)$
facilities and $k$ facilities for the residual instance becomes small (actually $0$ in this example).

Regarding the proof of Theorem~\ref{theorem:pseudo-approximation}, we first use
Theorem~\ref{theorem:construct-bi-point-solution} to obtain a bi-point
solution for $k$-median whose cost is at most twice the optimum cost. Jain and Vazirani~\cite{JV01} showed how to
convert a bi-point solution to an integral solution, losing a multiplicative factor of $2$ in the
approximation.  As we previously mentioned, this factor of $2$ is tight, as the fractional solution for the gap instance in
Figure~\ref{fig:gap-2-instance} is a bi-point solution.  Thus, this approach can only yield a
$4$-approximation.

This is where the $c$-additive pseudo-approximation is used and again the integrality gap instance
depicted in Figure~\ref{fig:gap-2-instance} inspired our approach. Recall that if we open the $k+1$
leaves of that instance, then we get a solution of cost $0$. In other words, by opening $1$
additional facility, we can do better than the fractional solution.  One may argue that this trick
is too weak to handle more sophisticated cases and try to enhance the gap instance. A natural way to
enhance it is to make many separate copies of the instance to obtain several ``stars''. One might
expect that the fractional cost in each copy is 1, the integral cost in each copy is 2 and opening
$1$ more facility can only improve the integral solution of one copy and thus does not improve the
overall ratio by too much. However, the integral solution can do much better since one cannot
restrict the integral solution to open $k$ facilities in each star.  As an example, consider the
case where we have $2$ copies. The integral solution can open $k-1$ facilities in the first star,
and $k+1$ facility in the second star.  Then, the cost of this solution is $3$, as opposed to $4$
achieved by opening $k$ facilities in each star. The gap is already reduced to $1.5$, without
opening additional facilities. Thus, this simple way to enhance the instance failed. 

Our pseudo-approximation algorithm is based on this intuition. From the bi-point solution $a \mF_1 +
b\mF_2$, we obtain copies of ``stars'' (similar to the integrality gap instance). Then for each star
we (basically) open either its center with probability $a$ or all its leaves with probability
$b$. Note that since either the center or all leaves of a star is open we have that a client always
has a ``close'' facility opened. With this intuition we prove in
Section~\ref{section:pseudo-approximation} that the expected cost of the obtained pseudo-solution is at
most $\frac{1+\sqrt{3}+\epsilon}{2}$ times the cost of the bi-fractional solution if we open
$O(1/\epsilon)$ additional facilities. The $O(1/\epsilon)$ additional facilities (and the case
distinction in Section~\ref{section:pseudo-approximation}) comes from the difficulty of handling
stars of different sizes. If all stars are of the same size the pseudo-approximation algorithm
becomes easier (run the algorithm in Section~\ref{sec:difficultalgocase} with one group of stars) and one obtains a
$\frac{1+\sqrt{3}}{2}$-approximate solution that opens at most $k+3$ facilities.


\section{Obtain solutions from additive pseudo-solutions}
\label{section:transformation}

In this section, we prove Theorem~\ref{theorem:transformation}. 
 As we mentioned earlier, there are
instances where pseudo-solutions opening $k+1$ facilities may have much smaller cost than solutions opening $k$ facilities. A
key concept to overcome this issue is the notion of sparse instances:
\begin{definition}
\label{def:sparseinstance}
For $A > 0$,  an  instance $\mI = (k, \mF,  \mC, d)$ is $A$-sparse if for each facility $i \in \mF$,
\begin{align}
\label{eq:sparscond}
(1-\xi) d(i, \OPT_\mI) \cdot |\CBall_\mI(i, \xi d(i, \OPT_\mI))| &\leq A,
\end{align} 
where $\xi := 1/3$.
We shall also say
that a facility $i$ is $A$-dense if it violates~\eqref{eq:sparscond}.
\end{definition}
\noindent Recall that $d(i, \OPT_\mI)$ is the distance from $i$ to its nearest facility in $\OPT_\mI$. 

The idea of the above definition is to avoid instances where we can significantly reduce the cost by
opening $O(1)$ additional facilities.  Consider the gap instance $\mI$ in
Figure~\ref{fig:gap-2-instance} and suppose $\OPT_\mI$ opens the center and the first $k-1$
leaf-facilities.  Then $\mI$ is not $A$-sparse for $A < \opt_\mI/2$ since
the last two leaf-facilities are $A$-dense.

The usefulness of the definition is twofold. On the
one hand, we show that we can concentrate on very sparse instances without loss of generality. On
the other hand, we show that any $c$-additive pseudo-solution to a sparse instance can be turned into a solution that
opens $k$ facilities by only increasing the cost slightly. The intuition behind the result that we
can only concentrate on sparse instances is the following. Consider an instance $\mI$ that is not
$\opt_\mI/t$-sparse for some constant $t$. If we consider a facility $i$ that is $\opt_\mI/t$-dense then the connection cost of the clients contained in $\CBall(i, \xi
d(i,\OPT_\mI))$ in the optimal solution $\OPT_\mI$ is at least $(1-\xi)d(i,\OPT_\mI) |\CBall(i, \xi
d(i, \OPT_\mI))| > \opt_\mI/t$. So, there can essentially (assuming disjointedness of the balls
of clients)
only be a constant $t$ number of facilities that violate the sparsity condition. We can guess this
set of dense facilities, as well as their nearest facility in $\OPT_\mI$ in time $n^{O(t)}$.

This is the intuition of
Algorithm~\ref{algo:preprocess} (that tries to guess and remove $\opt/t$-dense facilities) and the proof of the following lemma which is given in
Section~\ref{sec:proofsparseenum}. 

\ifdefined \stoc
\begin{algorithm*}[ht!]
\caption{Obtaining a solution from a $c$-additive pseudo-solution.}
\label{algo:transform}
 \algsetup{indent=1cm}
\begin{algorithmic}[1]
\REQUIRE{an $A$-sparse instance $\mI = (k, \mF, \mC, d)$, a
  $c$-additive pseudo-solution $\mT$, an integer $t\geq c$ and $\delta \in (0, 1/8)$}\;
\ENSURE{A solution $\mS$ satisfying the properties of Lemma~\ref{lemma:transform} }\;\\[2mm]
\STATE $\mT' := \mT$ and $B: = 2\cdot \frac{A + \cost_{\mI}(\mT)/t}{\delta\xi}$\;
\STATE \textbf{while} $\card{\mT'} > k$ and there is a facility $i \in \mT' $ such that $\cost_{\mI}(\mT' \setminus \set{i}) \leq \cost_{\mI}(\mT') + B$ \textbf{do}\label{STATE:start-reduce-T}\;
\STATE \hspace{\algorithmicindent}  Remove $i$ from $\mT'$;\label{STATE:end-reduce-T}
\RETURN $\mS:=\mT'$ if $\card{\mT'} \leq k$\label{STATE:firstreturn}; \\[2mm]
\STATE \textbf{for all} $\mD \subseteq \mT'$ and $\mV \subseteq \mF $ such that $|\mD|
+ |\mV| = k $ and $|\mV| < t$ \textbf{do} \label{STATE:guessing}\;\\[1mm]
\STATE \hspace{\algorithmicindent} For  $i \in \mD$, let $L_i = d(i, \mT'\setminus \set{i})$ and $f_i$ be the facility in $\FBall(i, \delta L_i)$ that minimizes
$$\sum_{j \in \CBall(i, L_i/3)} \min\set{d(f_i, j), d(j, \mV)}$$\label{STATE:define f-i}
\STATE \hspace{\algorithmicindent} Let $\mS_{\mD, \mV} := \mV \cup \{f_i : i \in \mD\}$ \;\\[1mm]
\RETURN $\mS :=\arg \min_{S_{\mD, \mV}} \cost_\mI(S_{\mD,\mV})$\;
\end{algorithmic}
\end{algorithm*}
\fi

\begin{lemma}
\label{lemma:sparseenum}
Given a $k$-median instance $\mI = (k,\mF, \mC, d)$ and a positive integer $t$,
Algorithm~\ref{algo:preprocess} outputs in time $n^{O(t)}$ many $k$-median instances obtained by removing facilities from $\mI$ so that at least one, say $\mI' = (k, \mF' \subseteq \mF, \mC, d)$, satisfies
\begin{enumerate}[(\ref{lemma:sparseenum}a):]
\item the optimal solution $\OPT_\mI$ to $\mI$ is also an optimal solution to $\mI'$; and
\item $\mI'$ is $\opt_\mI/t$-sparse.
\end{enumerate}
\end{lemma}

Note that $\mI'$ is obtained by removing facilities from $\mI$. Therefore any solution to $\mI'$ defines a
solution to $\mI$ of the same cost and  we can thus restrict our attention to sparse instances. 
The next lemma shows the advantage of
considering such instances. Assume we now have a $c$-additive solution $\mT$ to a  sparse instance $\mI$. 
Algorithm~\ref{algo:transform} tries
first in Lines~\ref{STATE:start-reduce-T}-\ref{STATE:end-reduce-T} to identify facilities in $\mT$ whose removal does not increase the cost by
too much. If the removal results in a set of at most $k$ facilities, we have obtained a ``good'' solution returned at Step~\ref{STATE:firstreturn} of the algorithm. Otherwise, as we prove  in
Section~\ref{sec:transform} using sparsity, more than $k-t$ of the facilities of the solution $\mT$ 
are very close to facilities in $\OPT_\mI$. Algorithm~\ref{algo:transform} therefore tries to guess
these facilities (the set $\mD$) and the remaining facilities of $\OPT_{\mI}$ (the set $\mV$). The
obtained bounds are given in the following lemma.
\begin{lemma}
\label{lemma:transform}
Given an $A$-sparse instance $\mI = (k, \mF, \mC, d)$, a
$c$-additive pseudo-solution $\mT$, $\delta \in (0,1/8)$, and  an integer $t\geq 2c/(\delta \xi)$,
Algorithm~\ref{algo:transform} finds in time $n^{O(t)}$ a set $\mS \subseteq \mF$ such
that:
\begin{enumerate}[(\ref{lemma:transform}a):]
\item\label{transform:a} $\mS$ is a solution to $\mI$, i.e, $|\mS| \leq k$; and
\item\label{transform:b} $\cost_\mI(\mS) \leq \max \left\{\cost_\mI(\mT) + cB,
    \frac{1+3\delta}{1-3\delta} \cdot \opt_\mI\right\}$, where $B := 2\cdot \frac{A +
    \cost_{\mI}(\mT)/t}{\xi \delta}$.
\end{enumerate}
\end{lemma}

Before giving the proofs of Lemmas~\ref{lemma:sparseenum} and~\ref{lemma:transform} let us see how
they imply the main result of this section.
\ifdefined \stoc
\paragraph{Proof of Theorem~\ref{theorem:transformation}} 
\else
\paragraph{Proof of Theorem~\ref{theorem:transformation}.} 
\fi
Select the largest $\delta\in (0,1/8)$ such that $(1+3\delta)/(1-3\delta) \leq \alpha$ and $t:= \frac{4}{\epsilon} \cdot \frac{\alpha c}{\xi \cdot \delta}
= O(c/\epsilon)$. Given a $k$-median instance $\mI$, use  Algorithm~\ref{algo:preprocess}
to obtain a set of $k$-median instances such
that at least one of these instances, say $\mI'$, satisfies the properties of
Lemma~\ref{lemma:sparseenum}. In particular, $\mI'$ is $\opt_\mI/t$-sparse. Now use algorithm $\mA$ to obtain $c$-additive pseudo-solutions to each of these
instances. Note that when we apply $\mA$ to $\mI'$, we obtain a solution $\mT$ such that
$\cost_{\mI}(\mT) = \cost_{\mI'}(\mT) \leq \alpha \cdot \opt_{\mI'} = \alpha \cdot \opt_\mI$. Finally, use Algorithm~\ref{algo:transform} (with the
same $t$ and  $\delta$ selected as above)  to transform the pseudo-solutions into real solutions and return the
solution to $\mI$ of minimum cost. The cost of the returned solution is at most the cost of $\mS$ where $\mS$ is the solution
obtained by transforming $\mT$. By Lemmas~\ref{lemma:sparseenum} and~\ref{lemma:transform}, we have that $\cost_{\mI}(S) = \cost_{\mI'}(S)$ is at most
$$
\max\left\{ \cost_{\mI} (\mT) + c \cdot 2\frac{\opt_{\mI}+ \cost_{\mI}(\mT)}{t
  \xi \delta}, \frac{1+3\delta}{1-3\delta} \opt_{\mI}\right\},
$$
which in turn, by the
selection of $\delta, \xi$,  and $t$, is at most $ \alpha \opt_{\mI} + c \cdot \frac{4 \alpha
  \opt_\mI}{t\xi \delta} \leq (\alpha + \epsilon) \opt_\mI$.

We conclude the proof of Theorem~\ref{theorem:transformation} by observing that the 
runtime of the algorithm is $n^{O(t)} = n^{O(c/\epsilon)}$ times the runtime of $\mA$. 

\subsection{Proof of Lemma~\ref{lemma:sparseenum}: obtaining a  sparse instance}
\label{sec:proofsparseenum}

\ifdefined \stoc
\else
\begin{algorithm}[t]
\caption{Enumeration of $k$-median instances.}
\label{algo:preprocess}
 \algsetup{indent=1cm}
\begin{algorithmic}
\REQUIRE{a $k$-median instance $\mI = (k, \mF, \mC, d)$ and a positive integer $t$}\;
\ENSURE{a set of $k$-median instances so that at least one satisfies the properties of Lemma~\ref{lemma:sparseenum}}\;\\[2mm]
\STATE \textbf{for all} $t'\leq t$
facility-pairs $(i_1, i_1'), (i_2, i_2'), \dots, (i_{t'}, i_{t'}')$ \textbf{output} $(k, \mF',
\mC, d)$, where \; \\[4mm]
\STATE \hspace{\algorithmicindent} $\mF' = \mF \setminus \bigcup_{z=1}^{t'}\FBall(i_z, d(i_z,i'_z))$
\COMMENT{{\scriptsize the facilities that are  closer to $i_z$ than $i'_z$ is to $i_z$ are removed }}\;\\[2mm]
\end{algorithmic}
\end{algorithm}
\fi

First note that Algorithm~\ref{algo:preprocess} selects $n^{O(t)}$ facility-pairs and can be
implemented to run in time $n^{O(t)}$.  We proceed by showing that for one selection of
facility-pairs the obtained instance satisfies the properties of Lemma~\ref{lemma:sparseenum}. Consider a maximal-length sequence $(i_1, i'_1), (i_2, i'_2), \dots, (i_\ell, i'_\ell)$ of facility-pairs satisfying: for every $b= 1, \dots, \ell$, 
\begin{itemize}
\item $i_{b} \in \mF \setminus  \bigcup_{z=1}^{b-1}\FBall(i_z, d(i_z,i'_z))$ is an $\opt_\mI/t$-dense
facility; and
\item $i'_{b}$ is the closest facility to $i_{b}$ in
$\OPT_{\mI}$.
\end{itemize}

Note that the instance $\mI' := (k, \mF', \mC, d) $ with $\mF' =  \mF \setminus
\bigcup_{z=1}^{\ell}\FBall(i_z, d(i_z,i'_z))$ is $\opt_\mI/t$-sparse  since
otherwise the sequence $(i_1, i'_1), (i_2,
i'_2), \dots, (i_\ell, i'_\ell)$ would not be of maximal length. Moreover, since we do not remove
any facilities in $\OPT_\mI$, i.e., $(\mF \setminus \mF') \cap \OPT_\mI = \emptyset$, we have that
$\OPT_\mI$ is also an optimal solution to $\mI'$.  In other words, $\mI'$  satisfies the
properties of  Lemma~\ref{lemma:sparseenum}. 

We complete the proof by showing that
Algorithm~\ref{algo:preprocess} enumerates $\mI'$, i.e., that $\ell \leq t$.  For the sake of
notation let $\mB_z := \CBall(i, \xi d( i_z, i'_z))$. First,  note that the client-balls $\mB_1, \mB_2, \dots, \mB_\ell$ are
disjoint. Indeed, if a ball $\mB_z$ overlaps a ball $\mB_w$ with $1\leq z < w\leq \ell$ then $d(i_z,i_w) < \xi d(i_z,
i'_z) + \xi d(i_w, i'_w)$. However,  since  $i_w$ must be in $\mF \setminus \FBall(i_z, d(i_z,i'_z))$, we have $d(i_z,i_w) \geq
d(i_z, i'_z)$. Since $i'_w$ is the closest facility in $\OPT_\mI$ to $i_w$, we have $d(i_w, i'_w) \leq d(i_w, i'_z)$, which, by triangle inequalities, is at most $d(i_z, i_w) + d(i_z,
i'_z) \leq 2 d(i_z, i_w)$. Hence (using that $\xi = 1/3$),
\begin{align*}
\xi(d( i_z, i'_z) + d(i_w, i'_w))         & \leq 3 \xi d(i_z, i_w) \leq d(i_z, i_w),
\end{align*}
which implies that the balls do not overlap. 

Second, note that the connection cost of a
client in $\mB_z$ is, by triangle inequalities, at least $(1-\xi) d(i_z, i_z') = (1-\xi) d(i_z,
\OPT_\mI)$.  We thus have (using that the client-balls are disjoint) that $\opt_\mI \geq  \sum_{z=1}^\ell (1-\xi) d(i_z,
\OPT_\mI) |\mB_z|$. As we only selected $\opt_\mI/t$-dense
facilities, $(1-\xi) d(i_z,
\OPT_\mI) |\mB_z| \geq \opt_\mI /t$ and hence
$\opt_\mI \geq \ell \opt_\mI/t$. It follows that $t \geq \ell$ which completes the proof of Lemma~\ref{lemma:sparseenum}.

\ifdefined \stoc
\newpage
\fi
\subsection{Proof of Lemma~\ref{lemma:transform}: obtain solution to sparse instance from pseudo-solution}
\label{sec:transform}
\ifdefined \stoc
\else
\begin{algorithm}[t]
\caption{Obtaining a solution from a $c$-additive pseudo-solution. }
\label{algo:transform}
 \algsetup{indent=1cm}
\begin{algorithmic}[1]
\REQUIRE{an $A$-sparse instance $\mI = (k, \mF, \mC, d)$, a
  $c$-additive pseudo-solution $\mT$, an integer $t\geq c$ and $\delta \in (0, 1/8)$}\;
\ENSURE{A solution $\mS$ satisfying the properties of Lemma~\ref{lemma:transform} }\;\\[2mm]
\STATE $\mT' := \mT$ and $B: = 2\cdot\frac{A + \cost_{\mI}(\mT)/t}{\delta\xi}$\;
\STATE \textbf{while} $\card{\mT'} > k$ and there is a facility $i \in \mT' $ such that $\cost_{\mI}(\mT' \setminus \set{i}) \leq \cost_{\mI}(\mT') + B$ \textbf{do}\label{STATE:start-reduce-T}\;
\STATE \hspace{\algorithmicindent}  Remove $i$ from $\mT'$;\label{STATE:end-reduce-T}
\RETURN $\mS:=\mT'$ if $\card{\mT'} \leq k$\label{STATE:firstreturn}; \\[2mm]
\STATE \textbf{for all} $\mD \subseteq \mT'$ and $\mV \subseteq \mF $ such that $|\mD|
+ |\mV| = k $ and $|\mV| < t$ \textbf{do} \label{STATE:guessing}\;\\[1mm]
\STATE \hspace{\algorithmicindent} For  $i \in \mD$, let $L_i = d(i, \mT'\setminus \set{i})$ and $f_i$ be the facility in $\FBall(i, \delta L_i)$ that minimizes
$$\sum_{j \in \CBall(i, L_i/3)} \min\set{d(f_i, j), d(j, \mV)}$$\label{STATE:define f-i}
\STATE \hspace{\algorithmicindent} Let $\mS_{\mD, \mV} := \mV \cup \{f_i : i \in \mD\}$ \;\\[1mm]
\RETURN $\mS :=\arg \min_{S_{\mD, \mV}} \cost_\mI(S_{\mD,\mV})$\;
\end{algorithmic}
\end{algorithm}
\fi

We start by analyzing the running time of Algorithm~\ref{algo:transform}. Clearly the while loop can
run at most $c$ iterations (a constant).
The number of different pairs $(\mD, \mV)$ in the for loop is at most
$$
\sum_{\ell = 0}^t   {|\mT'|  \choose k-\ell}{ |\mF| \choose \ell} .$$ 
Notice that $\card{\mT'} \leq k + c$ and $c \leq t$.  For sufficiently large $k$ and $\card{\mF}$, the above quantity is at most ${ |\mF| \choose t}\sum_{\ell
  = 0}^t  {k+c \choose c+\ell}  = n^{O(t)}$.
Algorithm~\ref{algo:transform} can thus be implemented to run in time $n^{O(t)}$ as
required. Moreover, it is clear from its definition that it always returns a solution $\mS$, i.e., $|\mS| \leq k$. 

We proceed by proving that $\mS$ satisfies (\ref{lemma:transform}b) of Lemma~\ref{lemma:transform}. Suppose first that the algorithm returns at
Line~\ref{STATE:firstreturn}.  By the condition of the while loop from
Line~\ref{STATE:start-reduce-T} to \ref{STATE:end-reduce-T}, we increase $\cost_{\mI}(\mT')$ by at
most $B$ each time we remove an element from $\mT'$. We remove at most $c$ elements and thus we
increase the total cost by at most $cB$. It follows that (\ref{lemma:transform}b) is immediately satisfied in this case.

From now on suppose instead that we reached Line~\ref{STATE:guessing} of
Algorithm~\ref{algo:transform} and thus $\card{\mT'} > k$.
We shall exhibit sets $\mD_0$ and $\mV_0$ such that $|\mD_0| + |\mV_0| = k$, $|\mV_0| < t$
and $\cost(\mS_{\mD_0, \mV_0}) \leq \frac{1+3\delta}{1-3\delta} \opt_\mI$. As Algorithm~\ref{algo:transform} selects $\mD_0$
and $\mV_0$ in one iteration and it returns the minimum cost solution, this concludes the proof of
Lemma~\ref{lemma:transform}. In order to define the sets $\mD_0$ and $\mV_0$ it shall be convenient to use the following definitions. 

\begin{definition}
For every facility $i \in \mT'$, let $L_i = d(i, \mT' \setminus \set{i})$ be the distance from $i$
  to its nearest neighbor in $\mT'$, and let $\ell_i = d(i, \OPT_\mI)$  be the minimum distance from $i$ to
  any facility in $\OPT_\mI$.

For a facility $i \in \mT'$, we say $i$ is \emph{determined} if $\ell_i < \delta L_i$. Otherwise, we say $i$ is \emph{undetermined}. 
\end{definition}

The sets $\mD_0$ and $\mV_0$ are now defined as follows. Set $\mD_0$ contain all facilities in $i\in
\mT'$ that are determined. If we let $f_i^*$ for $i\in \mD_0$ be the facility in
$\OPT_\mI$ that is closest to $i$, then set $\mV_0 := \OPT_\mI  \setminus \{f_i^*: i \in
\mD_0\}$. The intuition of  $\mD_0$ and $\mV_0$ is that the solution $\mS_{\mD_0, \mV_0}$ is very close to $\OPT_\mI$: the only
difference is the selection of $f_i$ at Line~\ref{STATE:define f-i} of Algorithm~\ref{algo:transform} instead of $f_i^*$. Since each $i\in \mD_0$ is determined, 
selecting $f_i$ greedily using a ``locally'' optimal strategy gives a good solution.  

\begin{figure}[ht]
\centering
\resizebox{\imagewidth}{!}{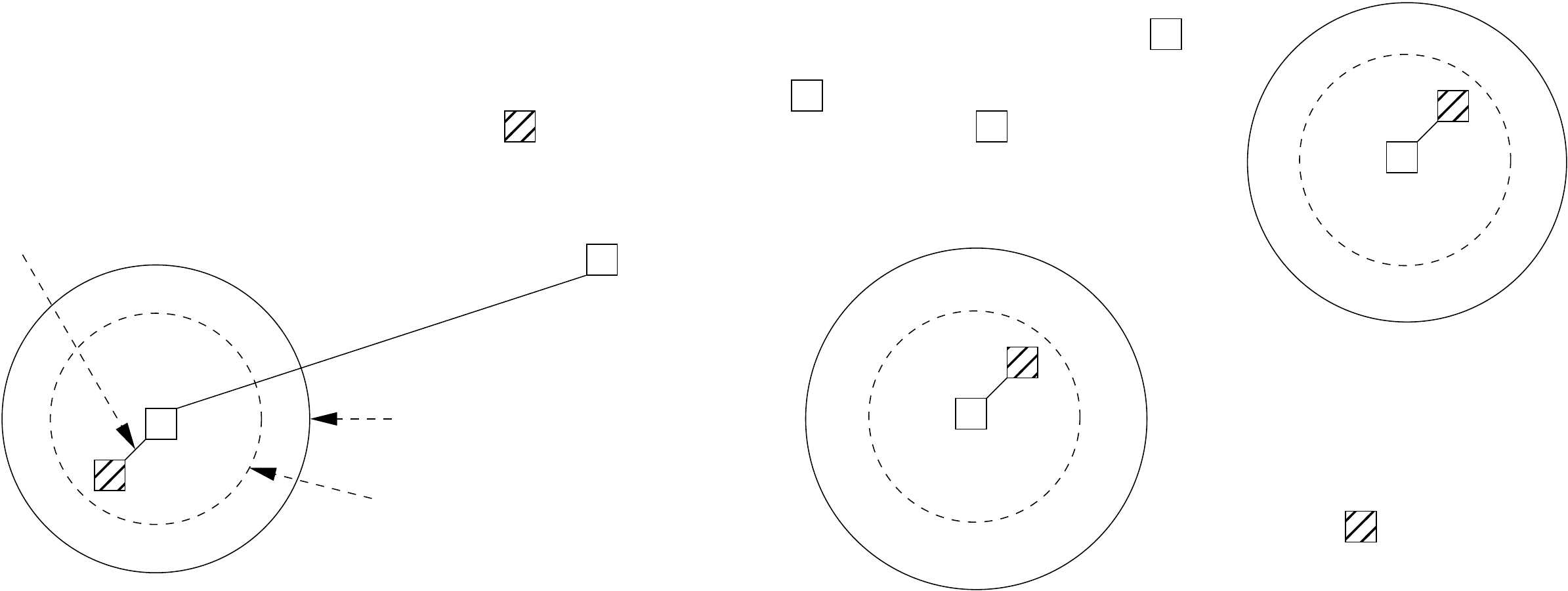}
\caption{Definitions of $\mD_0, \mV_0$ and $\mU_0$. Dashed and empty squares represent facilities in $\OPT_\mI$ and $\mT'$  respectively. $\mD_0$ is the set of empty squares circles. A dashed circle represents $\FBall(i, \delta L_i)$ for a determined facility $i \in \mD_0$. Thus, $f^*_i$ is in the ball since $\ell_i < \delta L_i$. $\mU_0$($\mV_0$, resp.) is the sets of empty (dashed, resp.) squares that are not inside any circle.  A solid circle for $i \in \mD_0$ represents the ``care-set'' of $i$.}
\label{fig:fac-types}
\end{figure}

We first show that sets $\mD_0$ and $\mV_0$ are indeed selected by Algorithm~\ref{algo:transform}
and then we conclude the proof of the lemma by bounding the cost of $\mS_{\mD_0, \mV_0}$.

\begin{claim}
$|\mD_0| + |\mV_0| = k$ and $|\mV_0| < t$.
\end{claim}
\begin{proofclaim} 
  We start by proving that $|\mD_0| + |\mV_0| = k $.  Recall that $\mV_0 = \OPT_\mI \setminus
  \{f_i^*: i \in \mD_0\}$. It is not hard to see that $f^*_i \neq f^*_{i'}$ for two distinct
  facilities in $\mD_0$.  This is indeed true since $d(i, i') \geq \max(L_i, L_{i'}), d(i, f^*_i)
  \leq \delta L_i, d(i', f^*_{i'}) \leq \delta L_{i'}$ and $\delta\leq 1/8$. Thus, $f^*(\mD_0) := \set{f^*_i : i \in \mD_0}$
  has size $\card{\mD_0}$, which in turn implies that (to simplify calculations we assume
  w.l.o.g. that $|\OPT_\mI| = k$)
$$
|\mV_0| = |\OPT_\mI| -  |\mD_0| = k - |\mD_0|.
$$

We proceed by proving $|\mV_0| < t$.  Note that the sets of determined and undetermined
facilities partition $\mT'$. Therefore, if we let $\mU_0$ be the set of undetermined facilities, we
have that $ |\mD_0| = |\mT'| - |\mU_0| .  $ Combining this with the above expression for
$|\mV_0|$ gives us
$$
|\mV_0| = k - |\mT'| + |\mU_0| \leq |\mU_0|.
$$ 
We complete the proof of the claim by showing that $|\mU_0| < t$. 

By the assumption that we reached Line~\ref{STATE:guessing} of Algorithm~\ref{algo:transform}, we
have $\card{\mT'} > k$ and $\cost_{\mI}(\mT' \setminus \set{i}) > \cost_{\mI}(\mT') + B$ for every $i
\in \mT'$.  Assume towards contradiction that $\card{\mU_0} \geq t$.  For every $i \in
\mT'$, let $\mC_i$ be the set of clients in $\mC$ connected to $i$ in the solution $\mT'$ and $C_i$ be
the total connection cost of these clients.  Thus, $\cost_{\mI}(\mT') = \sum_{i \in \mT'}C_i$. Take
the facility $i \in \mU_0$ with the minimum $C_i$. Then, we have $C_i \leq \cost_{\mI}(\mT')/t$. Let
$i'$ be the nearest neighbor of $i$ in $\mT'$; thus $d(i, i') = L_i$.

We shall remove the facility $i$ from $\mT'$ and connect the clients in $\mC_i$ to $i'$.  In order
to consider incremental connection cost incurred by the operation, we divide $\mC_i$ into two
parts.
\begin{description}
\item[$\mC_i \cap \CBall(i, \delta\xi L_i)$.] Since $i$ is undetermined, we have $\delta L_i \leq
  \ell_i$ and $\CBall(i, \delta\xi L_i) \subseteq \CBall(i, \xi \ell_i)$. 
  As $\mI$ is an $A$-sparse instance, $i$ is not an $A$-dense
  facility. That is $(1-\xi)\card{\CBall(i, \xi\ell_i)}\ell_i \leq A$, implying
\[
(1+\delta\xi)\card{\mC_i \cap \CBall(i, \delta\xi L_i)}L_i \leq \frac{(1+\delta\xi)}{\delta(1-\xi)}A
\leq A/(\delta \xi).
\] 
Then, as each client in $\mC_i \cap \CBall(i, \delta\xi L_i)$ has distance at most
$(1+\delta\xi)L_i$ to $i'$ (by triangle inequalities), connecting all clients in $\mC_i \cap
\CBall(i, \delta \xi L_i)$ to $i'$ can cost at most $
A/(\delta \xi)$.
\item[$\mC_i \setminus \CBall(i, \delta\xi L_i)$.] Consider any client $j$ in this set. Since $d(j,
  i') \leq d(j, i) + L_i$ and $d(j, i) \geq \delta\xi L_i$, we have $\frac{d(j, i') - d(j,
    i)}{d(j,i)} \leq \frac{L_i}{\delta\xi L_i} = 1/(\delta \xi)$.  Hence, the connection cost of a
  single client is increased by at most a factor $1/(\delta \xi)$. Therefore, the total connection
  cost increases by at most $C_i/(\delta \xi)$, which by the selection of $i$ is at most
  $\cost_{\mI}(\mT')/(\delta \xi t).$
\end{description}

Summing up the two quantities, removing $i$ from $\mT'$ can only increase the connection cost by at
most $ \frac{A + \cost_{\mI}(\mT')/t}{\delta \xi}$. As the while loop  of
Algorithm~\ref{algo:transform} ran for less than $c$ iterations, $\cost_{\mI}(\mT') <
\cost_{\mI}(\mT) + cB$. Therefore, $\frac{A + \cost_{\mI}(\mT')/t}{\delta \xi}  < \frac{A +
  (\cost_{\mI}(\mT) + cB)/t}{\delta \xi}$ which since $t \geq 2c/( \delta \xi)$ is at most 
\ifdefined \stoc
\ \\
\fi $\frac{A +
  \cost_{\mI}(\mT)/t}{\delta \xi} + B/2 = B$  leading to a
contradiction. Hence, $|\mU_0|< t$ which concludes the proof of the claim.
\end{proofclaim}

Having proved that the instance $\mS_{\mD_0, \mV_0}$ is selected by Algorithm~\ref{algo:transform},
we conclude the proof of Lemma~\ref{lemma:transform} by bounding the cost of $\mS_{\mD_0, \mV_0}$.

Recall that, for every $i \in \mD_0$, Algorithm~\ref{algo:transform} opens one facility $f_i $ in
the ball $\FBall(i, \delta L_i)$. We know we can do this so that the connection cost of $\mC$ is
$\opt_\mI$. We show that we can approximate this instance within a factor of $1 + O(\delta)$.
Roughly speaking, if a client is far away from any of these balls, then it does not care which
facilities to open inside the balls, up to a factor $1 + O(\delta)$.  If a client is close to one of
these balls, say $\FBall(i, \delta L_i)$, then we put the client into the ``care-set'' of $i$.  For
each $i$, we open a facility in the ball that is best for its care-set.

To be more specific, let the care-set of $i$ be $\CBall(i, L_i/3)$ for any $i \in \mD_0$.  Clearly, the balls $\CBall(i, L_i/3), i \in \mD_0$ are disjoint.  As
stated in Line~\ref{STATE:define f-i} of Algorithm~\ref{algo:transform}, we open a facility $f_i$ in
$\FBall(i, \delta L_i)$ that minimizes
\[
\sum_{j \in \CBall(i, L_i/3)}\min\set{d(f_i, j), d(j, \mV_0)}.
\] 
\begin{claim}
\label{claim:cost-small-2}
$\cost_{\mI}(\mS_{\mD_0, \mV_0}) \leq \frac{1+3\delta}{1-3\delta}\opt_\mI$.
\end{claim}
\begin{proofclaim}
  We compare $\OPT_\mI$ and $\mS_{\mD_0, \mV_0}$.  Consider a client $j \in \CBall(i, L_i/3)$ for some $i \in \mD_0$. The
  distance from $j$ to any facility in $\FBall(i, \delta L_i)$ is at most $(1/3 + \delta)L_i$. For
  any distinct facility $i' \in \mD_0$, the distance from $j$ to any facility in $\FBall(i', \delta
  L_{i'})$ is at least $d(i, i') - L_i/3 - \delta L_{i'} \geq d(i, i') - d(i, i')/3 - \delta d(i,
  i') = (2/3 - \delta)d(i,i') \geq (2/3 - \delta)L_i$. For $\delta \leq 1/8$, $1/3+\delta <
  2/3-\delta$. Thus, $j$ is either connected to $f^*_i$ or some facility in $ \mV_0$ in the
  solution $\OPT_\mI$.  Noticing that we are selecting the best $f_i$ for every $i \in \mD_0$, the total
  connection cost of $\bigcup_{i \in \mD_0}\CBall(i, L_i/3)$ in the solution $\mS_{\mD_0, \mV_0}$ is at most that in
  $\OPT_\mI$.

  Now, consider a client $j$ that is not in $\bigcup_{i \in \mD_0}\CBall(i, L_i/3)$. If it is
  connected to some facility in $ \mV_0$ in the solution $\OPT_\mI$, then the connection cost of
  $j$ in the solution $\mS_{\mD_0, \mV_0}$ can not be larger, since $ \mV_0 \subseteq \mS$.  Assume $j$ is
  connected to $f^*_i \in \CBall(i, L_i/3)$ for some $i \in \mD_0$. We compare $d(j, f^*_i)$ to $d(j,
  f_i)$:
\[
\frac{d(j, f_i)}{d(j, f^*_i)} \leq \frac{d(j, i) + \delta L_i}{d(j, i) - \delta L_i} \leq \frac{L_i/3 + \delta L_i}{L_i/3 - \delta L_i} = \frac{1 + 3\delta}{1-3\delta}.
\]

Thus, $\mS_{\mD_0, \mV_0}$ has connection cost at most $\frac{1+3\delta}{1-3\delta}\opt_\mI$.
\end{proofclaim}

\section{An $O(1/\epsilon)$-additive $1 + \sqrt{3} + \epsilon$ approximation for $k$-median}
\label{section:pseudo-approximation}

This section is dedicated to prove Theorem~\ref{theorem:pseudo-approximation}. 
Given a $k$-median instance $\mI = (k, \mF, \mC, d)$, we first use
Theorem~\ref{theorem:construct-bi-point-solution} to obtain a bi-point solution $a\mS_1 + b\mS_2$
whose cost is at most $2$ times the optimum cost of $\mI$.  Then it suffices to convert $a\mS_1 +
b\mS_2$ into an $O(1/\epsilon)$-additive solution, whose cost is at most
$\frac{1+\sqrt{3}+\epsilon}{2}$ times that of $a\mS_1 + b\mS_2$.

\begin{figure}[hbtp]
\center
\includegraphics[width=\imagewidth]{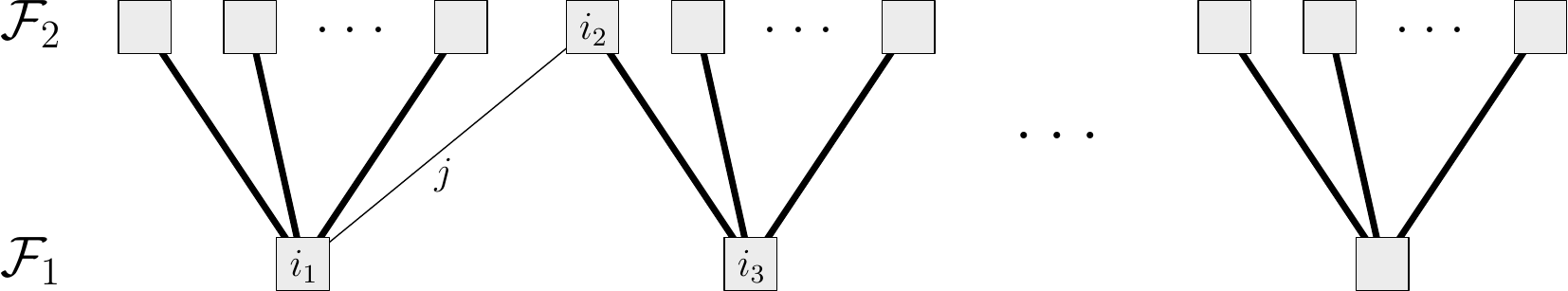}
\caption{Depiction of the bipartite graph associated to a bi-point solution. The fat edges are the
clients that form the edges of the stars. For clarity, we only depicted one client $j$ that is not part
of a star. Client $j$ has distances $d(j,i_1) = d_1(j), d(j,i_2) = d_2$ and $d(j,i_3)\leq  2d_2(j) +d_1(j)$.   }
\label{fig:stars}
\end{figure}

By the definition of bi-point solutions, we have $a + b = 1, |\mF_1| \leq k < |\mF_2|$ and  $a |\mF_1| +b|\mF_2| = k$. It shall be convenient to
think of $a\mF_1 + b\mF_2$ as a bipartite graph (see Figure~\ref{fig:stars}) with vertex sets
$\mF_1$ and $\mF_2$ and an edge for each client $j\in \mC$ that is incident to its closest
facilities in $\mF_1$ and $\mF_2$ denoted by $i_1(j)$ and $i_2(j)$, respectively. Moreover, let
$d_1(j) := d(j, i_1(j))$ and $d_2(j) := d(j, i_2(j))$.  Then, the (fractional) connection cost of $j$ in the
bi-point solution is $ad_1(j) + bd_2(j)$.  Similarly, if we let $d_1 := \cost(\mF_1) = \sum_{j \in \mC}d_1(j)$ and
$d_2 :=\cost(\mF_2) = \sum_{j \in \mC}d_2(j)$ then the bi-point solution has cost $ad_1 + bd_2$.

We shall prove Theorem~\ref{theorem:pseudo-approximation} by exhibiting different algorithms based
on the value of $a$. Specifically,  we shall distinguish between the cases when $a$ is in $\left(0, \frac{\sqrt{3}-1}{4}\right], \left(\frac{\sqrt{3}-1}{4}, \frac{2}{1+\sqrt{3}}\right],$
and $\left(\frac{2}{1+\sqrt{3}}, 1\right]$. The simplest case is when $a\in
\left(\frac{2}{1+\sqrt{3}}, 1\right]$: the solution where we open all facilities in $\mF_1$ is then a
$\frac{d_1}{ad_1 + b d_2} \leq 1/a = (1+\sqrt{3})/2$ approximation.

For the two remaining cases, we will use the concept of \emph{stars}. For each facility $i \in
\mF_2$ define $\pi(i)$ to be the facility in $\mF_1$ that is closest to $i$.  For a facility $i \in
\mF_1$ , let $S_i = \{i' \in \mF_2 : \pi(i') = i\}$. We think of $S_i$ as a star with \emph{center} $i$ and
\emph{leaves} $S_i$.  Note that by the definition of stars, we have that any client $j$ with $i_2(j) \in
S_i$ has $ d(i_2(j), i) \leq d(i_2(j), i_1(j)) = d_2(j) + d_1(j)$ and therefore $d(j,i) \leq d(j,
i_2(j)) + d(i_2(j), i) \leq 2d_2 + d_1$.  Our algorithms will ensure that there is an open facility
``close'' to every client by always opening $i$ if not all facilities in $S_i$ are opened.
The strategy for either opening the center of a star or its leaves (or sometimes both) depends on
the value of $a$. We start in Section~\ref{sec:simplealgocase} by explaining the simpler case when $a\in \left(0,
  \frac{\sqrt{3}-1}{4}\right]$ and then complete the proof of
Theorem~\ref{theorem:pseudo-approximation} by considering the final case in Section~\ref{sec:difficultalgocase}.

\subsection{Algorithm for $a\in \left(0,
  \frac{\sqrt{3}-1}{4}\right]$}
\label{sec:simplealgocase}
The idea behind our algorithm is that when $a$ is small then we can open most facilities in
$\mF_2$. We shall do so by starting with the trivial solution $\mF_1$ that we will improve by
almost greedily selecting stars and open all their leaves while closing their centers. 

As we will maintain the property that $i$ is open if not all
facilities in $S_i$ are open, we have that the connection cost of a client $j$ is 
$d_2(j)$ if $i_2(j)$ is open and at most $d_1(j) + 2 d_2(j)$ otherwise.  
%
Consider the trivial solution where we open all facilities in $\mF_1$. Then the total connection cost
is upper-bounded by $\sum_{j}(d_1(j) + 2d_2(j))$. If we open the facilities in $S_i$ instead of $i$
this will save us the cost $\sum_{j\in \delta(S_i)} (d_1(j) + d_2(j))$, where $\delta(S_i)$ denotes
the clients that are incident to the facilities in $S_i$. This motivates the following linear
program that maximizes the cost we will save compared to the trivial solution:
$$
\max \sum_{i\in \mF_1} \sum_{j \in \delta(S_i)} (d_1(j) + d_2(j)) x_i \qquad \mbox{ subject to}
$$
\begin{align*}
\sum_{i\in \mF_1} x_i (|S_i|-1)& \leq k - \card{\mF_1}  \\
0\leq x_i & \leq 1, \quad \forall i \in \mF_1
\end{align*}
Intuitively, $x_i$ takes value $1$ if we open all the facilities in $S_i$ and $0$ if we open $i$.
If $x_i = 1$, we need to open $\card{S_i} - 1$ more facilities (i.e, close $i$ and open all
facilities in $S_i$). Thus, the constraint says that we can only open $k$ facilities.  Note that
this is a Knapsack LP and hence it is easy to see that an optimal solution has at most one
fractional variable. Furthermore, $x_i = b$ is a feasible solution since $b \card{\mF_2} -
b\card{\mF_1} = k - \card{\mF_1}$. Therefore, the optimal solution to the LP has value at least
$b(d_1 + d_2)$.

Consider an optimal solution to the Knapsack LP with at most 1 fractional variable. Then, we open
all the facilities in $S_i$ with $x_i = 1$,
all the facilities $i \in \mF_1$ with $x_i = 0$, and
for the $i$ with fractional
$x_i$ we open $i$ and $\ceil{x_i|S_i|}$ facilities in $S_i$ uniformly at random. (This step can
easily be derandomized by greedily selecting the $\ceil{x_i|S_i|}$ facilities in $S_i$ that maximizes the reduced cost.)

Note that we opened the facilities so that the (expected) saved cost compared to the trivial solution is at least the value of
the optimal solution to the linear program. Therefore, 
this gives us a solution of (expected) cost at most $2d_2 + d_1 - b(d_2 + d_1 )= (1+a) d_2 + a d_1$. Also, the
solution opens at most $k+2$ facilities, where the additive term 2 comes from the star $S_i$ with
fractional $x_i$ value.

Since we can assume that $d_2 \leq d_1$ (otherwise we can simply open all facilities in $\mF_1$),
the algorithm has an approximation guarantee of
$$
\frac{(1+a) d_2  + ad_1}
{(1-a) d_2 + a d_1} \leq (1+ 2a),
$$
which is at most $\frac{1+\sqrt{3}}{2}$ if $a \leq \frac{\sqrt{3}-1}{4}$.

\subsection{Algorithm for $a\in \left(\frac{\sqrt{3}-1}{4}, \frac{2}{1+\sqrt{3}}\right]$}
\label{sec:difficultalgocase}
In this subsection, we give the algorithm for the most complex case. To simplify the arguments, we
give a randomized algorithm that can easily be derandomized using the standard method of conditional
probabilities. The idea is that we wish to describe a randomized rounding that opens a facility in
$\mF_1$ with probability $\approx a$ and a
facility in $\mF_2$ with probability $\approx b$ and at the same time ensuring that there always is an open
facility ``close'' to a client by maintaining the property: if $i$ is not open then all facilities in
$S_i$ are open for all stars. 

We now describe such a randomized rounding that takes a parameter $\eta>0$ that balances the achieved
approximation guarantee with the amount of additional facilities we open: the achieved approximation
ratio is $(1+\eta)\frac{1+\sqrt{3}}{2}$ while we open at most $k+O(1/\eta)$ facilities. It shall
be convenient to distinguish between large and small stars. We say that a star $S_i$ is \emph{large}
if $|S_i| \geq 2/(ab\eta)$ and \emph{small} otherwise. Moreover, we partition the small stars into
$\ceil{2/(a b\eta)}$ groups according to their sizes:
\[
\mU_h = \{i\in \mF_1: |S_i| = h\} \qquad \mbox{for } h=0,1,\dots, \ceil{2/(ab\eta)}-1.
\]
The randomized algorithm can now be described as follows:
\begin{algorithmic}[1]
\STATE For each large star $S_i$: open $i$ and open $\floor{b(|S_i|-1)}$ facilities in $S_i$
  uniformly at random. \;
\STATE For each group $\mU_h$ of small stars: take a random permutation of the stars in $\mU_h$,
open the centers of the first $\ceil{a|\mU_h|} + 1$ stars, and open all leaves of the remaining
stars. In addition, if we let $L$ be the number of already opened leaves subtracted from $bh |\mU_h|$, then
with probability $\ceil{L}- L$ open $\floor{L}$ and with remaining probability open $\ceil{L}$ randomly picked leaves  in the first $\ceil{a|\mU_h|} + 1$ stars.
\end{algorithmic}
Note that for a large star the algorithm always opens its center and (almost) a $b$ fraction of its leaves. For a group $\mU_h$
of small stars, note that we open either the center (with probability at least $a$) or all leaves of a star. Moreover, we open the
additional leaves
 so that in expectation exactly a $b$ fraction of the leaves of the stars in $\mU_h$ are opened.

We start by showing that the algorithm does not open too many facilities; we then continue by bounding
the expected cost of the obtained solution. 
\begin{claim}
 The algorithm opens  at most $k+ 3\ceil{2/(ab\eta)}$
 facilities.
\end{claim}
\begin{proofclaim}
Recall that we have that $a\card{\mF_1} + b\card{\mF_2} = k$ and therefore
\begin{equation}
\label{eq:starsum}
\sum_{i\in \mF_1}\left(a + b \card{S_i}\right)= k.
\end{equation}

First, consider a large star $i\in \mF_1$, i.e., $a |S_i| \geq 1/(b\eta) \geq 1/\eta$.  For such a star,
the algorithm opens $1 + \lfloor b(|S_i| - 1)\rfloor \leq  1 + b(|S_i| -1) = a + b|S_i|$ facilities, which
is the contribution of star $i$ to~\eqref{eq:starsum}.

Second, consider a group $\mU_h$ of small stars and let $m:= |\mU_h|$. When considering this group,
the algorithm opens $\lceil am\rceil + 1 \leq am + 2$ facilities in $\mF_1$, and at most
\[
 (m - \ceil{am}- 1) h  + \ceil{ bh m -  (m - \ceil{am}- 1) h} \leq bhm + 1 
\]
facilities in $\mF_2$. Thus, the total number of facilities open from the group $\mU_h$ of small
stars is at most $m(a + bh) + 3$.  As $m$ is the size of $\mU_h$ and $a + bh$ is the
contribution of each star in $\mU_h$ to~\eqref{eq:starsum}, the statement follows from that we have at
most $\ceil{2/(ab\eta)}$ groups.
\end{proofclaim}

We proceed by bounding the expected cost of the obtained solution. The intuition behind the
following claim is that we have designed a randomized algorithm that opens a facility in $\mF_2$
with probability $\approx b$ and a facility in $\mF_1$ with probability $\approx a$. Therefore, if
we connect a client $j$ to $i_2(j)$ with connection cost $d_2(j)$ if that facility is open,  to
$i_1(j)$ with connection cost $d_1(j)$ if that facility but not $i_2(j)$ is open, and to the center $i$
of the star $S_i: i_2(j) \in S_i$ with connection cost at most $2d_2(j) + d_1(j)$ if neither
$i_1(j)$ or $i_2(j)$ are opened (recall that $i$ is open if not all facilities in $S_i$ are open), then the expected connection cost of client $j$ is  at most
\ifdefined \stoc
\begin{align*}
&\ \ \ b \cdot d_2(j) + (1-b) a \cdot d_1(j) + ab( 2d_2(j) + d_1(j)) \\
&= a d_1(j) + b(1+ 2a) d_2(j).
\end{align*}
\else
$$
b \cdot d_2(j) + (1-b) a \cdot d_1(j) + ab( 2d_2(j) + d_1(j)) = a d_1(j) + b(1+ 2a) d_2(j).
$$
\fi
The following claim then follows by linearity of expectation.

\begin{claim}
The algorithm returns a solution with expected cost at most
$$
(1+\eta) \left(a d_1 + b(1+2a) d_2 \right).
$$
\end{claim}
\begin{proofclaim}

  Focus on a client $j$ with $i_1(j) = i_1$ and $i_2(j) = i_2$ as depicted in Figure~\ref{fig:stars}. Let $i_3 = \pi(i_2)$ be the closest
  facility in $\mF_1$ to $i_2$, i.e., $i_3$ is the center of the star $S_{i_3}$ with $i_2 \in
  S_{i_3}$.   Notice that $d(i_3, i_2) \leq d(i_1, i_2) \leq d_1(j) + d_2(j)$ by the definition of $\pi$. Thus, $d(j, i_3) \leq d_2(j) + d(i_3, i_2) \leq d_1(j) + 2d_2(j)$. We connect $j$ to $i_2$, if $i_2$ is open; otherwise, we connect $j$ to $i_1$ if $i_1$ is open. We connect $j$ to $i_3$ if both $i_1$
  and $i_2$ are not open. (Notice that for a star $S_i$, if $i$ is not open, then all facilities in
  $S_i$ are open. Thus, either $i_2$ or $i_3$ is open.) Connecting $j$ to the nearest open facility can only give smaller connection cost.  By abusing notations we let $i_1$ ($i_2$, resp.) denote the event that $i_1$ ($i_2$, resp.) is open and $\overline i_1$ ($\overline i_2$, resp.) denote the event that $i_1$ ($i_2$, resp.) is not open. Then, we can upper bound the expected connection cost of $j$ by
$$
\Pr[i_2] \cdot d_2(j) + \Pr\left[i_1 \overline i_2\right] \cdot d_1(j) +
\Pr\left[\overline i_1 \overline i_2\right] \cdot (2d_2(j) + d_1(j)),
$$
which, by substituting  $\Pr\left[\overline i_1 \overline i_2\right]  = 1- \Pr[i_2] - \Pr\left[ i_1
  \overline i_2\right]$, equals
\begin{equation}
\label{eq:upper1}
\left(2- \Pr\left[i_2\right] - 2\Pr\left[i_1 \overline i_2\right]\right) d_2(j)
+ \left(1 - \Pr\left[i_2\right]\right) d_1(j).
\end{equation}
We upper bound this expression by analyzing these probabilities. 

Let us start with $\Pr\left[i_1 \overline i_2\right]$. If $i_2 \in S_{i_1}$ (i.e., $i_1 = i_3$) then $i_1$ is always open if $i_2$ is closed and thus we have $\Pr\left[i_1 \overline i_2\right] = \Pr\left[\overline i_2\right]$.   If $S_{i_1}$  is a
large star, then $i_1$ is always open and we also have $\Pr\left[i_1 \overline i_2\right] = \Pr\left[\overline i_2\right]$.  In both cases, we have $\Pr\left[i_1\overline i_2\right] = 1 - \Pr[i_2]$.

We now consider the case where $S_{i_1}$ is a small star in a group $\mU_h$ with $m:=|\mU_h|$ and $i_1 \neq i_3$. Note that if $S_{i_3}$ is either a large star or a small star not in $\mU_h$ then the events
$i_1$ and $\bar i_2$ are independent. We have thus in this case
that
\begin{align*}
\Pr\left[i_1 \overline i_2\right] &= \Pr[i_1] \cdot (1-\Pr[i_2]) \\
& = \frac{\ceil{am}+1}{m} \cdot (1-\Pr[i_2])
\end{align*}
It remains to consider the case when $S_{i_3}$ is a star in $\mU_h$. Notice that the
dependence between $i_1$ and $i_2$ comes from that if $i_2$ is closed then $i_3$ is
opened. Therefore, we have
\begin{align*}
\Pr\left[i_1 \overline i_2\right] &= \Pr\left[i_1| \overline i_2\right] \cdot (1-\Pr[i_2]) \\
& = \frac{\ceil{am}+1 -1}{m} \cdot (1-\Pr[i_2]).
\end{align*}

We have thus showed that $\Pr\left[i_1 \overline i_2\right]$ is always at least
$a\cdot (1- \Pr\left[i_2\right])$. Substituting in this bound in~\eqref{eq:upper1} allows
us to upper bound the connection cost of $j$ by
$$
\left(2b + (2a-1)\Pr\left[i_2\right]\right)d_2(j) + \left(1-\Pr\left[i_2\right]\right) d_1(j).
$$
We proceed by analyzing $\Pr\left[i_2\right]$. On the one hand, if $i_2$ is a leaf of some big star $S_i$ with $s = \card{S_i} \geq 2/(ba\eta)$ then 
$
\Pr[i_2] =  \frac{\floor{b(s -1)}}{s}$ is greater than $b-2/s \geq b(1-a\eta)$ and smaller than
$b$.  On the other hand, if $i_2$ is a leaf  of a small star $S_i$ in group $\mU_h$ with $m:= |\mU_h|$ then
in expectation we open exactly a $b$  fraction of the leaves so 
$
\Pr\left[i_2\right] = b$.
We have thus that $b(1-a\eta) \leq \Pr\left[i_2\right] \leq b$. Since $(1+\eta)\cdot (1-a\eta)
\geq 1$ we have that the expected connection cost of facility $j$ is at most $(1+\eta)$ times
$$
(2b+ (2a-1) b) d_2(j) + (1-b)d_1(j)  = b(1+ 2a) d_2(j) + ad_1(j).
$$

The claim now follows by summing up the expected connection cost of all clients.
\end{proofclaim}

We complete the analysis by balancing the solution obtained by running our algorithm with the
trivial solution of cost $d_1$ that opens all facilities in $\mF_1$.

\begin{claim}
We have that
$\min\set{d_1, ad_1+b(1+2a)d_2} \leq \frac{1+\sqrt{3}}{2}(ad_1 + bd_2)$.
\end{claim}
\begin{proofclaim}
We change $d_1$ and $d_2$ slightly so that $ad_1 + bd_2$ does not change. Apply the operation to the direction that increases the left-hand-side of the inequality.  This operation can be applied until one of the 3 conditions is true: (1) $d_1 = 0$; (2) $d_2 = 0$ or (3) $d_1 = ad_1 + b(1 + 2a)d_2$.

For the first two cases, the inequality holds. In the third case, we have $d_1 = (1 + 2a)d_2$. Then
$\frac{d_1}{ad_1 + bd_2} = \frac{1+2a}{a(1+2a) + 1-a} = \frac{1+2a}{1+2a^2}$. The maximum value of the quantity is $\frac{1 + \sqrt{3}}{2}$, achieved when $a = \frac{\sqrt{3}-1}{2}$.
\end{proofclaim}

We have shown that, by letting  $\eta =
\epsilon/(1+\sqrt{3})$, we can efficiently obtain a
$O(1/\epsilon)$-additive $\frac{1+\sqrt{3} +\epsilon}{2}$-approximation to a bi-point solution with
constant $a$ and $b$, which proves Theorem~\ref{theorem:pseudo-approximation} when $a \in  \left(\frac{\sqrt{3}-1}{4}, \frac{2}{1+\sqrt{3}}\right]$.

\section{Discussion}
We have given a $1+\sqrt{3}+\epsilon$-approximation algorithm for $k$-median, improving upon the
previous best $3+\epsilon$-approximation algorithm.  Besides the improved approximation guarantee,
we believe that the most interesting technical contribution is Theorem~\ref{theorem:transformation},
namely that we can approximate $k$ in $k$-median without loss of generality. More specifically, any
pseudo-approximation algorithm which outputs a solution that opens $k+O(1)$ facilities can be turned
into an approximation algorithm with essentially the same approximation guarantee but that only
opens $k$ facilities.

For $k$-median this new point of view has the potential to overcome a known barrier for obtaining an approximation algorithm
that matches the $1+2/e$ hardness of approximation result: the lower bound of $2$ on the integrality gap
of the natural LP for $k$-median.  In particular, the known instances that give the integrality gap of $2$
vanish if we allow $k+1$ open facilities in the integral solution. Following our work, we therefore find
it important to further understand the following open question: what is the maximum ratio between the cost of the optimum solution with $k+O(1)$ open
facilities, and the value of the LP with $k$ open facilities? One can note that the hardness of
approximation reduction in~\cite{JMS02} implies that the integrality gap is at least $1+2/e$ even if we
open $k+o(k)$ facilities. Moreover our $O(1/\epsilon)$-additive approximation for bi-point solutions
achieving a guarantee of $\frac{1+\sqrt{3}+\epsilon}{2} < 1+2/e$ shows that the worst case
integrality gap instances are not of this type when pseudo-approximation is allowed.

Finally, we would like to mention that Theorem~\ref{theorem:transformation} naturally motivates the
question if other hard constraints can be relaxed to soft constraints with a ``violation-dependent''
increase in the runtime. Soft constraints often greatly help when designing algorithms. For example,
the capacitated versions of facility location and $k$-median are notorious problems when the
capacities are hard constraints but better approximation algorithms are known if the capacities are
allowed to be slightly violated (see e.g.~\cite{ChuzhoyR05}). As our approach was inspired by studying the power of the
Sherali-Adams hierarchy~\cite{SheraliA90} for the $k$-median problem, we believe that a promising research direction is to understand the power
of that hierarchy and the stronger Lasserre hierarchy~\cite{Lasserre02} when applied to these kinds
of problems.

\end{document}